\documentclass[12pt]{article}
 \usepackage{amsfonts}
 \usepackage{amssymb}
 \newfont{\bbbold}{msbm10}

 \def\bbC{\mbox{\bbbold C}}

 \def\cA{{\cal A}}
 
 \def\cC{{\cal C}}
 \def\cD{{\cal D}}

 \def\cJ{{\cal J}}
 
 \def\cL{{\cal L}}
 \def\cM{{\cal M}}

 \def\\{{\cal R}}
 \def\cS{{\cal S}}
 \def\cT{{\cal T}}

 \newfont{\goth}{eufm10 scaled \magstep1}

 \def\gi{\mbox{\goth i}}

 \def\gn{\mbox{\goth n}}
 
 \def\gp{\mbox{\goth p}}

 \def\gs{\mbox{\goth s}}

 \def\a{\alpha}
 \def\b{\beta}
 \def\c{\gamma}\def\C{\Gamma}
 \def\d{\delta}
 
 \def\f{\phi}
 \def\h{\eta}
 \def\k{\kappa}
 \def\L{\Lambda}
 \def\m{\mu}

 \def\th{\theta}

 \def\be{\begin{equation}}\def\ee{\end{equation}}
 \def\bea{\begin{eqnarray}}\def\eea{\end{eqnarray}}
 \def\ba{\begin{array}}\def\ea{\end{array}}

 \def\del{\partial}
 \def\ua{\underline{\alpha}}
 \def\ub{\underline{\phantom{\alpha}}\!\!\!\beta}

 \def\una{\underline a}\def\unA{\underline A}
 \def\unB{\underline B}
 \def\unc{\underline c}\def\unC{\underline C}

 \def\unm{\underline m}\def\unM{\underline M}
 \def\unn{\underline n}

 \def\nab{\nabla}

 \def\del{\partial}

 \let\la=\label

 {}

 \def\nn{\nonumber}
 \def\bd{\begin{document}}
 \def\ed{\end{document}}
 \def\bea{\begin{eqnarray}}\def\barr{\begin{array}}\def\earr{\end{array}}
 \def\eea{\end{eqnarray}}
 \def\ft#1#2{{\textstyle{{\scriptstyle #1}\over {\scriptstyle #2}}}}
 \def\fft#1#2{{#1 \over #2}}
 \newcommand{\eq}[1]{(\ref{#1})}
 \def\eqs#1#2{(\ref{#1}-\ref{#2})}
 \def\det{{\rm det\,}}
 \def\tr{{\rm tr}}\def\Tr{{\rm Tr}}
	\def\slask{\hspace{0.em}\not\hspace{-0.15em}}

\begin{document}

 \thispagestyle{empty}

 \vspace{20pt}

 \begin{center}
 {\Large{\bf Higher Order Invariants in Supergravity}}\footnote{Invited lecture at ``Supergravity at
25'', Stony Brook, December 2001.}
 \vspace{60pt}

{U. Lindstr\"om}\footnote{e-mail: ul@physto.se} \vskip
.2cm{Department of Physics}, \linebreak
{Stockholm University},
SCFAB\\
{S-106 91 Stockholm, Sweden}\\
\vspace {15pt}

 \vspace{90pt}

 \end{center}

 {\bf Abstract}

On a historical note, we first describe the early superspace construction of counterterms in
supergravity and then move on to a brief discussion of selected areas in string theory where higher
order supergravity invariants enter the effective theories. Motivated by this description we argue
that it is important to understand $p$-brane actions with $\k$-invariant higher order terms, thus
re-opening the question of $\k$-invariant ``rigidity'' terms. Finally we describe a recent
construction of such an action using the superembedding formalism.

 \vfill\leftline{}\vfill \vskip  10pt

 \baselineskip=15pt \pagebreak \setcounter{page}{1}

\section{Introduction}

Twenty-two years ago, at the Supergravity workshop in Stony
Brook 1979 \cite{VanNieuwenhuizen:1979hm}, I gave one of my very
first talks in front of an international audience. A few
minutes into that talk the chairman had to break up a heated
discussion between some members of the audience, using the
phrase ``Gentlemen, gentlemen, duels will have to wait until
after the talk. Pistols will be provided at the back of the
room''. A nerve-racking first encounter with the
supergravity community.

So what was I talking about and why did it stir up the emotions?
The title of my talk was``Use of dimensional reduction in the
search for supergravity invariants''\cite{Lindstrom:1979gh},
and it was concerned with finding on-shell invariants that
might serve as counter terms in a quantization of
supergravity. The reason that this was a ``hot'' subject, you
recall, was that supergravity, or at least its maximally
extended version, was believed to be finite. If one could prove
the absence of counterterms that would be a very important
result.

At the time, three loop counterterms had been shown to
exist in supergravity at the linearized level
\cite{Deser:1977nt} ($N=1$),\cite{Deser:1978br} ($N=2$), and at
the full non-linear level \cite{Kaku:1977rk}
\cite{Ferrara:1978wj} ($N=1$). I was hoping to adress the
all-important $N=8$ case by studying the
$N=1$ theory in $11$ dimensions and dimensionally reduce to
four. As a warm-up I had looked at six-dimensional
supersymmetric Yang-Mills theory with action
\bea
\int d^6x[-{1\over 4}F_{m
n}F^{mn}+i\bar\lambda
\slask\partial\lambda]~,
\eea
and constructed the
following on-shell higher order invariant \cite{Deser:1980sx}: 
\be\int
d^6x[\cT_{mn}\cT^{mn}+\cD_{mn}\cD^{mn}+{i\over
2}\bar\cJ_m\slask\partial\cJ^m-{3\over
4}\cC_m\partial^2\cC^m]~,
\ee
where the stress energy tensor $\cT_{mn}$, the
supercurrent $\cJ_m$, and the spin density $\cC_m$ are the
currents expected from knowledge of the fourdimensional
invariant.  $\cD_{mn}$ however contains a new identically
conserved two-form ${}^*(F\wedge F)$. This was a first sign
of a feature that arises abundantly when trying to construct
on-shell invariants in $11$ dimensional supergravity. This
complicates the calculations a lot, and one needs a good way
of organizing the calculation, i.e., one should go to
superspace.

\section{Supergravity Counterterms}

At the Stony Brook conference I met Paul Howe who, with Lars
Brink, had just constructed the $N=8$ on-shell superspace
supergravity in four dimensions \cite{Brink:1979nt}. We
gradually began looking at the problem of constructing higher
order invariants and presented the results at
the first Nuffield meeting the following summer (1980)
\cite{Howe:1980du}. Here is a brief summary of our results
\cite{Howe:1981th}:\\

Extended supergravity in superspace is described by a
superfield $W$, which is a scalar for $N\geq 4$ and has spin
$2-N/2$ for $N<4$. The superspace tangent space group is
$SL(2,\bbC)\otimes G$, where $G$ is (a subgroup of) $U(N)$. The
superfield $W$ also carries internal indices corresponding to
$G$, and may further transform in a representation of some
global symmetry group $G'$ (=$E_7$ for $N=8$). This description
is on-shell. 

The linearized form of the n-loop counterterm Lagrangians we
consider is 
\be
\cL^{(n)}=\kappa^{2(n-1)}\{R^{n+1}+susy~completion\}~,
\ee
where $R^{n+1}$ is shorthand for a $n$-fold product of the
Riemann tensor with suitably contracted indices. In superspace
this corresponds to
\be
S^{(n)}= \int d^4xd^{4N}\theta\cL^{(n)}(z) ~,
\label{3loop}
\ee
where we restrict ourselves to full superspace integrals
(over all $z=x,\theta$). For
$N\leq 3$, there is a full non-linear $3$-loop counterterm
respecting all symmetries\footnote{At the linearized level,
this term had been constructed for $N=1$
\cite{Ferrara:1978mv}. The corresponding non-linear
\underline{off-shell}
$N=1$ term was only recently constructed \cite{Moura:2001xx}.}:
\be
\cL^{(3)}(z)=\kappa^4EW^2\bar{W}^2~,
\ee
with $E$ the super-determinant of the super-vielbein and 
summation of indices is supressed. At the linearized level Renata Kallosh constucted a 
$3$-loop counterterm also for $N=8$, although neither the supersymmetry nor the $SU(8)$ is
manifest \cite{Kallosh:1981fi}. Its manifest version was given in superspace in \cite{Howe:1981xy},
and in harmonic superspace in \cite{Hartwell:1995rp}.

For $N\leq 4$
there are 
$N-1$ loop invariants simililar to (\ref{3loop}), constructed
from products of the fundamental superfield to the appropriate
power, but they do not respect the global $G'$ invariance of
the field equations. To find fully invariant actions we had to
go to $N$-loops. The Lagrangian is 
\be
\cL^{(N)}=\kappa^{2(N-1)}E\epsilon^{\alpha\beta}\epsilon^{\dot\alpha\dot\beta}
\chi_{\alpha a b c}\chi_{\beta
def}\bar\chi^{abc}_{\dot\alpha}\bar\chi^{def}_{\dot\beta}~,
\label{Nloop}
\ee
where the torsion\footnote{An underlined index denotes a pair of tangent
space indices, e.g., $\underline{\gamma}\equiv (c,\gamma$). This notation is use only
here, and in later sections
underlining will denote ambient space time indices.}
$T_{\underline{\alpha\beta}}^{~~\dot{\underline{\gamma}}}=2\epsilon_{\alpha\beta}\bar\chi^{abc\dot\gamma}$,
and $\chi$ occurs as first spinor derivative of the
 fundamental superfield ($a,b,..$ are $G$-indices). At the linearized
level, the spin-two content of (\ref{Nloop}) is a D'Alembertian to the
$N-3$ power sandwiched between the square of the Weyl spinor and the square
of its complex conjugate.

In particular our results show that there are possible counterterms even
in the maximally extended supergravity, and point towards the rebirth of 
string theory and the modern
interpretation of supergravity as an effective theory.

\section{String Theory Effective Actions}

In string theory, supergravities have their role as effective theories for
describing the massless sector of the field theory limit to lowest order in
$\alpha'$, as well as, in $11$ dimensions, the low energy limit of
$\cM$-theory. As effective theories, the supergravity actions are expected
to receive higher order derivative corrections. In the $\cM$-theory
context in particular, there is very active research aimed at finding the
correct modification of supergravity \cite{Howe:1997rf, Cederwall:2000ye, Peeters:2001qj,
Peeters:2001ub}.

Further,  the lowest
order dynamics of $D$-branes in such a background is given by an
effective action which is a sum of the Dirac-Born-Infeld (DBI) action
and a Wess-Zumino-Witten (WZW) action $\cS=\cS_{DBI}+\cS_{WZW}$. The higher
order corrections to these actions are partly known. E.g., the WZW action,
which describes the coupling of $Dp$-branes to the bulk Ramond-Ramond
fields\footnote{Here $C=C^{(0)}+C^{(1)+...+C^{(9)}}$ is a formal sum of
the Ramond $n$-forms, odd forms contributing in the IIA and even in the
IIB theory.}, is
\cite{Green:1997dd, Cheung:1998az, Minasian:1997mm};
\be
\cS_{WZW}=T_p\int_{M^{p+1}} C\wedge
tr_N\left(e^{2\pi\alpha'F}\right)\wedge\left({{\hat\cA}(4\pi^2\alpha'R_T)
\over{\hat\cA}(4\pi^2\alpha'R_N)}\right)^{1/2}~,
\label{WZW}
\ee
where $T_p$ is the tension of the $Dp$-brane and the trace is in the
fundamental representation. The square root of the Dirac `roof' genus has
an expansion in even powers of the curvature two-form,
\be
\sqrt{\hat\cA}=1-{1\over48}p_1(R)+\frac{1}{2560}p^2_1(R)
-\frac{1}{2880}p_2(R)+...~,
\ee
with $p_1$ and $p_2$ the first two Pontryagin classes. The tangential and
normal curvatures that occur in (\ref{WZW}) are given by the Gauss-Codazzi
relations according to
\bea
&&(R_T)_{mnrs}=R_{mnrs}+\delta_{p't'}(\Omega^(p')_{m[r}\Omega^{t'}_{s]n})\\
&&(R_N)_{mn}^{~~p't'}=-R^{p't'}_{~~mn}+g^{rs}(\Omega^(p')_{r[m}\Omega^{t'}_{n]s})~,
\label{curve}
\eea
where $\Omega$ is the second fundamental form, unprimed indices are worldvolume
indices and primed indices refer to the normal bundle.

Similarily, higher order corrections to the DBI action,
\be
\cS_{DBI}=T_p\int
d^{p+1}xe^\Phi\sqrt{-det[(G_{\unm\unn}+B_{\unm\unn})\partial_mX^{\unm}
\partial_nX^{\unn}+F_{mn}]}~,
\ee 
have been calulated. (Here  $\Phi$ is the
dilaton and $B_{\unm\unn}$ the antisymmetric NS-NS field, and underlined
indices refer to the ambient spece-time.)

To second order in
$\alpha'$,  $\partial F$-corrections were calculated in \cite{Andreev:1988cb},
and higher order such corrections were computed for the combined DBI-WZW
action\footnote{In a constant background.}  in
\cite{Wyllard:2001qe}. 

The higher  $\partial^2X$ derivative corrections were
treated to to second order in $\alpha'$ in \cite{Bachas:1999um,
Fotopoulos:2001pt}. In a  background where the $B$ and $F$ fields vanish, the
corrections are given by
\bea
\cS^{(2)}_{DBI}=cT_p\int
d^{p+1}xe^{-\Phi}\sqrt{-g}(\alpha')^2[(R_T)_{mnrs}(R_T)^{mnrs}
-2(R_T)_{mn}(R_T)^{mn}\\
-(R_N)_{mnp't'}(R_N)^{mnp't'}
+2\bar{R}_{p't'}\bar{R}^{p't'}]~,
\label{second}
\eea
where $g$ is the determinant of the induced metric, $c$ is a numerical constant
and
$\bar{R}$ is obtained by contracting tangent indices only on the curvature
tensor and adding a term quadratic in the second fundamental form. The result in
(\ref{second}) is unique up to an ambiguity involving the trace of the second
fundamental form
$\Omega^{p'~m}_m$.

Now, all the higher order terms for the $Dp$-branes described above concern the
purely bosonic theory. At the lowest
order in $\alpha'$ and in an arbitrary supergravity background, a
particular combination $\cS=\cS_{DBI}+\cS_{WZW}$ is needed for
$\kappa$ symmetry of the corresponding Green-Schwarz (GS) action
\cite{Cederwall:1997ri}. It is thus important for consistency to understand
how the above results can be extended in a $\kappa$-symmetric way.

As seen from  (\ref{curve}), (\ref{WZW}) and (\ref{second}), this entails
finding $\kappa$-symmetric brane actions with terms that involve the second
fundamental form.  Such terms are often (sloppily) referred to as ``extrinsic
curvature terms'' or ``rigidity'' terms, and we are thus led to re-open the quest
for
$\kappa$-symmetric brane actions with rigidity terms.

\section{Rigid Branes}

Partly motivated by Polyakov's suggestion that the QCD string
should be viewed as a string with extrinsic curvature terms
\cite{Polyakov:1986cs}, "rigid" strings and $p$-branes were quite
extensively studied in the 80's,
\cite{Kleinert:1986bk,Curtright:1986ed,Lindstrom:1987ps} and early
90's \cite{Polchinski:1992ty}. 
The string action that was suggested in \cite{Polyakov:1986cs} is
\be
\cS_R=\frac{1}{2\pi\alpha'}\int d^2x\sqrt{-g}[1+\m\Omega^2]~,
\la{poly}
\ee
where
\be
\Omega^2\equiv \Omega^{p'~m}_m\Omega^{p'~n}_n\sim
\eta_{\unm\unn}\partial_m\partial^mX^{\unm}\partial_n\partial^nX^{\unn}~,
\ee
and $\m$ is a dimensionful coupling constant. In attempts to include this
action, or its generalization, in the bosonic sector of a GS
action for a $p$-brane, one has to consider
$\kappa$ symmetry, i.e., the generalization to $p$-branes of Siegel's symmetry
for the superparticle \cite{Siegel:1983hh}. In the simple setting of a flat
space background, the action for a the 1-brane (string) reads
\be
\cS_{GS}=T_p\int
d^2x[\sqrt{-g}+\epsilon^{mn}\bar\th\slask\Pi_m\partial_n\th]~,
\label{pGS}
\ee
where $g_{mn}$ is the globally supersymmetrized induced metric
\bea
&&g_{mn}\equiv \eta_{\unm\unn}\Pi_m^{\unm}\Pi_n^{\unn}\\
&&\Pi_m^{\unm}\equiv \partial_mX^{\unm}-i\bar\th\C^{\unm}\partial_m\th~.
\eea
The action (\ref{pGS}) is invariant under the following symmetry with a local
parameter $\k$:
\be
\delta\th=\k\qquad \delta X^{\unm}=-i\bar\k\C^{\unm}\th~.
\ee
(Apart from $\k$ being local, the only formal difference from the
global supersymmetry is a sign-change in $\delta X$). The parameter $\k$
satisfies a projection relation 
\be
\k=P\k~,
\label{proj}
\ee where $P^2=P$.

The outcome of the attempts to construct rigid $p$-branes may be summarized as
follows: A
$\kappa$-symmetric GS type "rigid" string was proposed in
\cite{Curtright:1987mr}, and a generalization to higher $p$ was suggested in
\cite{Curtright:1987mh}.  The $\kappa$-symmetry was only shown (for
the string) to second order in the spinorial target space coordinate $\theta$, however, and a fully
$\kappa$-symmetric formulation was only found for the "rigid"
superparticle \cite{Ivanov:1991ub}, \cite{Ivanov:1992cb},
\cite{Gauntlett:1991dw}. (For the 2-brane, spinning, i.e., locally
worldvolume supersymmetric, formulations with extrinsic curvature terms
were found \cite{Lindstrom:1988fr}). 

It is interesting to note that the most
complete result, the
$\kappa$-symmetric description of  the "rigid" superparticle, is based on
a description where 
$\kappa$-symmetry is embedded in a local worldline superconformal symmetry, a
fact which points to the way we understand $\kappa$-symmetry today, namely as
defined in terms of the local supersymmetry of the worldsurface in the
superembedding approach to $p$-branes. This formalism is thus a
natural starting point for a renewed attempt at finding a rigid
$\kappa$-symmetric $p$-brane.

\section{Superembeddings}

The superembedding formalism was first used in the context of superparticles 
\cite{Volkov:1988vf, Sorokin:1989zi, Sorokin:1989nj}, and has been applied to
various other branes, for a review see
\cite{Sorokin:2000jx}\footnote{\cite{Gates:1986vk} is an earlier attempt to use
source and target superspaces.}. In \cite{Howe:1997mx, Howe:1997yn} the
formalism is extended to include branes with gauge fields on the world volume
and the $M5$ brane dynamics derived. Below follows a brief description of the
formalism.

We consider an embedding of one superspace $M$ into another $\unM$, given in
terms of the coordinates as $X^{\unM}(X^M)$. As above, underlined indices refer
to the ambient (super-)space and the bare indices to the world volume. Tangent
space indices are from the beginning of the alphabet and world indices from the
middle. Lower case latin letters denote bosoe and lower case Greek letters
denote fermi indices. Hence, e.g., the ambient superspace has tangent space
indices $\unA=(\una,\ua)$. Indices for the normal bundle are denoted by primes,
as in, e.g.,  $(a',\a')$.

The embedding matrix is defined as follows
\be
E_A^{\unA}\equiv E_A^M\partial_MX^{\unM}E_{\unM}^{\unA}~.
\ee
The basic geometric embedding condition which ensures that the odd tangen space
of the world volume is a subspace of the odd tangent space of the target
superspace is:
\be
E_\a^{\una}=0~.
\label{emb}
\ee
To explore the consequences of the embedding condition (\ref{emb}) one specifies
the geometry of $\unM$, parametrizes the embedding and studies the following
torison equation:
\be
2\nabla_{[A}E_{B]}^{\unC}+T_{AB}^CE_C^{\unC}=
(-)^{A(B+{\unB})}E_B^{\unB}E_A^{\unA}T_{\unA\unB}^{\unC}~.
\label{tor}
\ee
This relation is the pull-back of the equation defining the target space torsion
two-form. (The covariant derivative acts on all types of tensor indices.)
The analysis of (\ref{emb}) and (\ref{tor}) yields the induced supergeometry on
the embedded brane in terms of multiplets which are typically on-shell for large
number of supersymmetries. For $N\leq 16$ it can also be off-shell. the
equations of the component or GS formalism can be obtained by taking
the leading ($\th^\mu=0$) components of the equations that decribe the brane
multiplet. These equations are guaranteed to be $\k$-symmetric. We summarize
the argument:

Let $v^M$ be a worldvolume vector field generating infinitesimal
diffeomorphisms. Then
\be
\delta X^{\unM}=v^M\partial_MX^{\unM}~,
\ee
or
\be
\delta X^{\unA}\equiv \delta X^{\unM}E_{\unM}^{\unA}=v^A E_{\unA}~.
\ee
For the special case when $v^a=0$, this gives, using (\ref{emb}),
\bea
&&\delta X^{\una}=0\\
&&\delta X^{\ua}=v^\a E_\a^{\ua}~.
\eea
We recover the usual $\k$-symmetry relations by defining
\be
\k^{\ua} \equiv v^\a E_\a^{\ua}~,
\ee
and noting that it has to lie in the worldvolume subspace of the odd tangent
space of the target space. The latter condition means that there is a projection
operator $P$ such that $\k$ satisfies (c.f. (\ref{proj})):
\be
\k^{\ua}=\k^{\ub} P_{\ub}^{\ua}~.
\ee

The fact that the embedding formalism yields $\k$-invariant results makes it
eminently suitable when looking for a $\k$-symmetric rigid $p$-brane, provided
there is an off-shell description of the embedding.

\section{A $\k$-symmetic Rigid Membrane}

The simplest example of the kind of models we are looking for is found in the
superembedding of a membrane in flat four-dimensional superspace. In this
section we highlight a few of its features, a full description may be found in
\cite{Howe:2001wc}.

 For the embedding matrix we can take,
\bea E_{\a}{}^{\una}&=&0 \\ E_{\a}{}^{\ua} &=& u_{\a}{}^{\ua} +
i\d_{\a}{}^{\b'}h  u_{\b'}{}^{\ua}\ , \label{3.1} \eea
where both $\a$ and $\a'$ are $d=3$ spinor indices taking two
values. We also have
\bea
&&E_a{}^{\una}= u_a{}^{\una}\\
&&T_{\a\b}{}^c=-i(\c^c)_{\a\b}\\ 
&& E_a{}^{\una}=(1+h^2) u_a{}^{\una}~;
\la{rel1}
\eea
where
\be
 {\rm Spin}(1,3)\ni u=\left(\ba{l} u_{\a}{}^{\ua} \\
u_{\a'}{}^{\ua}\ea\right)
 \la{3.3}
 \ee
and where the corresponding element of $SO(1,3)$ is made up by
$u_a{}^{\una}$ and a normal component $u_{3}{}^{\una}$. The geometry
allows us to complement the dimension zero worldvolume torsion in (\ref{rel1})
with the rest of the standard off-shell $N=1, d=3$ supergravity torsion
constraints \cite{Brown:1979ma}
\be
 \ba{cccccc}
 T_{\a\b}{}^{\c} &=& 0 & T_{\a b}{}^{c} &=& 0\\
 &&&&&\\
 T_{ab}{}^c &=& 0 & T_{a\b}{}^{\c}&=&i(\c_a)_{\b}{}^{\c} S~.
 \ea
 \la{3.7}
 \ee
The field $S$ is later determined in
terms of the worldvolume multiplet so that the geometry is indeed
induced.

From this starting point, we may systematically analyze the consequences of
(\ref{tor}), order by order in (mass) dimension. Below only  those
definitions needded to interpret our final result for the action are given. 

One of the
fields that enter, $h$, has already been introduced above. Another, $X_{ab3}$, is
one of the components of the $\gs\gp\gi\gn(1,3)$
valued one-form $X=du u^{-1}$, and  $\L_a{}^{\a} u_{\a}{}^{\ua}$, fianlly is
defined by
\be
 E_a{}^{\ua}=\L_a{}^{\a} u_{\a}{}^{\ua} + \psi_{a}{}^{\a'}
 u_{\a'}{}^{\ua}~.
 \la{3.9}
 \ee

A systematic procedure for constructing actions in the superembedding formalsim
is presented in \cite{Howe:1998ts}. There it was used to rederive the
GS action for the our membrane in the form
\be
\cS_{GS}^0=T_2\int
d^3x\left[\sqrt{-g}\left(\frac{1-h^2}{1+h^2}\right)-\frac{1}{6}\epsilon^{mnp}C_{mnp}\right]|~,
\la{per}
\ee  
where $|$ denotes taking the lowest component of the superfield. Since we want to find an
extension of Polyakov's action (\ref{poly}), the action (\ref{per}) provides an already
$\k$-symmetric starting point, but we have to provide the higher order terms. From the second term
in (\ref{poly}), we see that the higher order Lagrangian has to have dimension $1$. Such terms
may be found by constructing terms quadratic in $\L_a{}^{\a}$
multiplied by arbitrary functions of $h$ (since $h$ has dimension zero).

There are two possible quadratic terms one can construct
from $\L$,

 \be
 L^{(1)}=-{i\over2}\L^{a\a}\L_{a\a}
 \la{5.1}
 \ee

and

 \be
 L^{(2)}=-{i\over2}(\c^{ab})^{\a\b}\L_{a\a}\L_{b\b}~.
 \la{5.2}
 \ee

The contributions to the GS action from these terms are (neglecting fermions)
\bea
 -i\nab^{\a}\nab_{\a}L^{(1)}&=& {1\over2}X_{ab3}X^{ab3} +
 {h^2(2+9h^2)\over2(1+3h^2)^2} -{2(1+2h^2 +3h^4)\over f^2}(\nab
 h)^2\la{5.3}\\
 -i\nab^{\a}\nab_{\a}L^{(2)}&=&{1\over2}X_{ab3}X^{ab3}-\left({1\over2}+
 {h^2(2+9h^2)\over(1+3h^2)^2}\right) X_3^2+ {4h^2(2+3h^2)\over
 f^2}(\nab h)^2
 \la{5.3b}
 \eea

A general linear combination of these two terms with $h$-dependent
coefficients will lead to a rather complicated bosonic Lagrangian. Using the freedom to multiply by
arbirtary fuctions of $h$, one may ensure that the $h$-equation remains algebraic (as in
(\ref{per})) above). It would be very complicated in general, though. 
A second possibility
is to take the linear combination $L^{(1)}+{1\over2}L^{(2)}$. In
this case there are still derivative $h$ terms but all the
coefficient functions simplify dramatically. We find

 \be
 L_o^{(1)}+{1\over2}L_o^{(2)}={3\over4}\hat X_{ab3}\hat X^{ab3}-{2\over
 f^2} (\nab h)^2
 \la{5.4}
 \ee

where $\hat X_{ab 3}:=X_{ab3}-1/3\h_{ab} X_3$ is traceless.
 Lagrangian for the sum of the GS and
higher derivative Lagrangian:

 \be
 L(x)=\sqrt{-\det g}\left({1-h^2\over 1+h^2} +{\b\over2}\hat
 \Omega^2-{4\b\over3f^2} (\nab h)^2\right)
 \la{5.5}
 \ee

where

 \be
 \Omega_{mn}{}^3=(\nab_m \del_n x^{\unc} )(u^{-1})_{\unc}{}^3 \nn \\
 \la{5.6}
 \ee

with $\hat \Omega$ being the corresponding traceless tensor. The metric
here is

 \be
 g_{mn}=\del_m x^{\una} \del_n x_{\una}
 \la{5.7}
 \ee

i.e. the standard induced metric for the associated bosonic
embedding. The tensor $\hat \Omega_{mn}{}^3$ is then the traceless
second fundamental form for this embedding.

If we now set $h=\tan \phi$  the Lagrangian becomes

 \be
 L(x)=\sqrt{-\det g}( {\b\over2}\hat
 \Omega^2-{4\b\over3} (\nab \f)^2 + \cos 2\f)
 \la{5.5b}
 \ee

Hence the equation of motion for the auxiliary field in this case
is simply the sine-Gordon equation,

 \be
 \nab^2\f -{3\over4\b}\sin 2\f=0
 \la{5.6b}
 \ee

\section{Summary}

Higher derivative supergravity terms were studied earlier as counter terms
for the quantum theory. In that context, the superspace eight-loop counterterm (\ref{Nloop})
remains the only example of an higher order invariant in $N=8$ supergravity which respects
all the symmetries of that theory. There have been arguments put forward for the existence of a
five-loop counterterm, though \cite{Bern:2000fm}, \cite{Deser:1999jz}.

Nowadays higher derivative supergravity terms arise, e.g., in string theory low energy effective
actions. Investigations of corrections to
$D$-brane actions yield ``extrinsic curvature'' terms in the bosonic sector of the higher order
Lagrangians. The full GS actions must thus contain
$\k$-symmetric completions of these terms. When so-called ``rigid'' strings and branes were
studied previously such a $\k$-symmetrization proved notoriously difficult. With the advent of the
superembedding formalism we now have a systematic way of constructing such models, at least for the
cases where the multiplets governing the embeddings are off-shell. A membrane in flat
four-dimensional superspace is such a case and has been worked out in detail.

An open question is if it is possible to extend the results to $D$-branes and verify that the
known bosonic sectors may be $\k$-symmetrized. Perhaps the superembedding formalism needs to be
modified to allow off-shell descriptions of the relevant embeddings.

\section*{Acknowledgements}

I would like to thank the organizers of ``Supergravity at 25'' for providing such a convincing
argument for the vitality of the subject. I am also grateful to Paul Howe and Bengt Nilsson for
valuable comments.

\pagebreak

\end{document}